\documentclass[12pt,preprint]{aastex}
\usepackage{graphics}
\input epsf
\begin{document}

\title{A Turbulent Origin for Flocculent Spiral Structure in Galaxies:
II. Observations and Models of M33\footnote{Based on data
obtained at the Canada-France-Hawaii Telescope}}

\author{Bruce G. Elmegreen \affil{IBM Research Division, T.J. Watson
Research Center, P.O. Box 218, Yorktown Heights, NY 10598, USA,
bge@watson.ibm.com} }

\author{Samuel N. Leitner \affil{Wesleyan University, Dept. of Physics, Middletown,
CT; sleitner@wesleyan.edu}}

\author{Debra Meloy Elmegreen \affil{Vassar College,
Dept. of Physics \& Astronomy, Box 745, Poughkeepsie, NY 12604;
elmegreen@vassar.edu} }

\author{Jean-Charles Cuillandre \affil{Canada-France-Hawaii Telescope,
65-1238 Mamalahoa Highway, Kamuela, HI 96743, jcc@cfht.hawaii.edu}}

\begin{abstract}
Fourier transform power spectra of azimuthal scans of the optical structure
of M33 are evaluated for B, V, and R passbands and fit to fractal models
of continuum emission with superposed star formation.  Power spectra
are also determined for H$\alpha$. The best models have intrinsic power
spectra with 1D slopes of around $-0.7\pm0.7$, significantly shallower
than the Kolmogorov spectrum (slope $=-1.7$) but steeper than pure noise
(slope $=0$).  A fit to the power spectrum of the flocculent galaxy
NGC 5055 gives a steeper slope of around $-1.5\pm0.2$, which could be
from turbulence.  Both cases model the optical light as a superposition
of continuous and point-like stellar sources that follow an underlying
fractal pattern.  Foreground bright stars are clipped in the images, but
they are so prominent in M33 that even their residual affects the power
spectrum, making it shallower than what is intrinsic to the galaxy.
A model consisting of random foreground stars added to the best model
of NGC 5055 fits the observed power spectrum of M33 as well as the
shallower intrinsic power spectrum that was made without foreground stars.
Thus the optical structure in M33 could result from turbulence too.

\end{abstract}

\section{Introduction}

The flocculent spiral structure in several nearby galaxies was
recently shown to have a power spectrum for azimuthal scans that
resembles the power spectrum of HI emission from the Large
Magellanic Clouds (Elmegreen, Elmegreen, \& Leitner 2003;
hereafter Paper I). This power spectrum has a long-wavelength part
that falls as $\sim 1/k$ for wavenumber $k$, an intermediate part
that falls approximately as $k^{-5/3}$, and a short wave part that
falls as $\sim k^{-1}$ again.  Individual stars contribute most to
the short waves, and the brightest stars can dominate this part of
the power spectrum if they are not removed.  The distance scale
for the $k^{-5/3}$ part typically extends up to several hundred
parsecs, depending on galaxy and galacto-centric radius. For NGC
5055, which is a large flocculent galaxy, the $\sim k^{-5/3}$ part
extends up to 1 kpc.

Aside from the short-wave contamination from individual bright stars,
these optical power spectra closely resemble the power spectrum of HI
emission from the LMC (Elmegreen, Kim, \& Staveley-Smith 2001) and dust
absorption from the nuclear regions of two Sa-type galaxies (Elmegreen,
Elmegreen \& Eberwein 2002).  This similarity led us to suggest that
young stars follow the turbulent gas as they form, and that the outer
scale for this turbulence is comparable to the disk thickness or the
inverse Jeans length in the interstellar medium.  The implication of
this is not only that gravitational instabilities form flocculent arms,
which was well-known before from numerical simulations (Sellwood \&
Carlberg 1984), but also that these instabilities generate much of the
turbulence in the interstellar medium, which both structures the gas
and causes the stars to form in fractal patterns (see also Huber \&
Pfenniger 2001; Wada, Meurer, \& Norman 2002).

Paper I reviewed the observations of power spectra in local gas and
dust (e.g., Crovisier \& Dickey 1983; Gautier, et al. 1992; Green 1993;
Armstrong et al. 1995; St\"utzki et al. 1998; Schlegel, Finkbeiner, \&
Davis 1998; Deshpande, Dwarakanath, \& Goss 2000; Dickey et al. 2001) and
in whole galaxies (Stanimirovic, et al. 1999; Stanimirovic et al. 2000;
Elmegreen, Kim, \& Staveley-Smith 2001). It also discussed the possible
links between gaseous density structures and the structures that come from
velocity in a turbulent medium (e.g., Lazarian \& Pogosyan 2000; Goldman
2000; Stanimirovic \& Lazarian 2001; Lazarian et al. 2001; Lithwick \&
Goldreich 2001; Cho \& Lazarian 2003).  Here we model the stellar light
distribution in two galaxies to illustrate that star formation follows
this turbulent gas and acquires a similar power spectrum.  The nature
of interstellar turbulence and star formation are not understood well
enough to explain the results.  Several processes are possible, including
the formation of clouds and stellar complexes in moving sub-clouds that
act like passive scalars in a large-scale turbulent flow (Goldman 2000;
Boldyrev, Nordlund, \& Padoan 2002; Paper I), the formation of clouds
and stars in turbulence-compressed regions (Klessen, Heitsch, \& Mac
Low 2000; Ossenkopf, Klessen, \& Heitsch 2001; Klessen 2001; Padoan et
al. 2001a; Padoan \& Nordlund 2002) and the formation of stars in shocks
that are driven by other stars in a turbulent medium (Elmegreen 2002a).
All of these processes give about the same power law structure for density
and star formation.

The advent of large-scale digital CCD images measuring several
thousand pixels on a side has made power spectrum analyses of
galactic structure meaningful.  Large images are necessary to get
enough dynamic range to see the power law part of a power spectrum
if there is one.  In this respect, the optical survey of galaxies
by the CFH12K camera of the Canada-France-Hawaii telescope (CFHT)
is ideal.  Here we analyze the CFHT image of
M33 (Cuillandre, Lequeux, \& Loinard 1999), which has an original
pixel size of 0.206 arcsec and a binned size of 0.412 arcsec for
the full galaxy mosaic. The maximum azimuthal circumference for
the part we use is 10000 px, or 4120 arcsec, which  gives scales
covering 4 orders of magnitude. We take power spectra in both the
azimuthal direction at equally spaced radii and along the spiral
direction, following the arm and interarm regions.

We also model this power spectrum and the power spectrum of NGC
5055, a flocculent galaxy from Paper I, with fake-galaxy images of
a fractal Brownian motion continuum plus discrete stars that
follow this continuum.

\section{Power Spectra of the Galaxy M33}

Figure 1 shows two images of M33 with azimuthal and spiral lines at
which scans of intensity were obtained and converted to 1D power spectra.
The azimuthal scans are circles in the galaxy disk plane and the spiral
scans go through both arm and interarm regions at a constant pitch
angle (24.5 degrees).  This pitch angle fits the strong arm in the
south using deprojected coordinates (see Sandage and Humphreys 1980
for pitch angle fits to each arm).  Power spectra were obtained for
intensity scans along these directions, with each scan one pixel wide
to maximize k-space resolution.  For the azimuthal scans, 8 adjacent
pixel-wide power spectra were averaged together to make the final power
spectrum at each radius. These 8 adjacent scans for the 10 selected
radii are shown in the figure as dark pixels. For the spiral scans,
21 adjacent pixel scans were used separately for each power spectrum,
and then these 21 power spectra were averaged together to give the
resultant power spectrum along each of the 8 spiral or interspiral cuts.

Figure 2 has the azimuthal intensity scan from the ellipse that is
fourth out from the center of the R-band image, and it shows the
power spectrum of this scan multiplied by $k^{5/3}$. The original
scan is on the left and a version with the brightest stars removed
is on the right. The power spectrum is the sum of the squares of
the sine and cosine Fourier transforms, plotted in log-log as a
function of the wavenumber, $k$.  The abscissa in the figure is
the wave number normalized so that the right-most point, with a
value of 1, has $k=1/2$ pixels$^{-1}$. The distance corresponding
to any normalized $k$ value is $2/k$ pixels. For our images one
pixel is 1.68 parsecs, assuming a distance to M33 of 0.84 Mpc
(Freedman, Wilson \& Madore 1991).

The power spectra in Figure 2 and other figures here have been
multiplied by $k^{5/3}$ in order to flatten what is normally a
$k^{-5/3}$ spectrum for incompressible Kolmogorov turbulence and
make this part easier to see. There are many possible explanations
for a power spectrum that is approximately $k^{-5/3}$, as reviewed
in detail in Paper I.  It is probably not the result of motions in
a perfectly incompressible fluid, as for the Kolmogorov problem,
but some degree of incompressibility is still possible (Goldman
2000; Boldyrev, Nordlund, \& Padoan 2002;
Lithwick \& Goldreich 2001).  Even if it is highly
compressible, the power spectrum of density structure is about the
same (see Lazarian \& Pogosyan 2000; Cho \& Lazarian 2003).

Bright stars are clipped from the intensity scans in 3 steps.
First we find the average intensity level of a clipped version of
the scan, clipped below zero and at a high enough level to remove
the brightest and saturated stars, all of which are foreground.
Then we find the running boxcar average 31 pixels wide of the same
scan, clipped a second time at some intermediate height above the
average to remove the pedestals of the brightest stars, but not
clipped low enough to remove the stars in M33.  Finally, we clip
the scan at a level above the running average that is 3 times the
Gaussian $\sigma$ for the rms noise in this final clipped version;
this last step is done with several iterations.  The first step is
necessary to account for the exponential disk; the second to
include large-scale variations from star-formation regions and
spiral waves, and the third to reduce the impact of stars in the
power spectrum.

The power spectrum has a slightly steeper rise in Figure 2 when
bright stars are included, and it has a deeper dip at high
frequency. M33 has many foreground stars and bright point sources
inside the galaxy so star removal is difficult. Thus there is a
residual high-$k$ dip in all of the results shown here.  The
rising part could also be a bit too steep as a result of residual
stars.  This translates into a non-normalized power spectrum that
is too shallow in its decline. The galaxies in Paper I had fewer
foreground stars and shallower rising parts in their
$k^{5/3}-$normalized power spectra.  One of them, NGC 5055, will
be modelled in the next section.

Power spectra of M33 for ten radii in three passbands are shown in
Figure 3, and power spectra for the spiral scans in R band, along
with the intensity profiles averaged over 21 adjacent single-pixel
wide scans (as discussed above), are shown in Figure 4.  The power
spectra are all similar regardless of color, radius, arm, or
interarm position. After multiplication by $k^{5/3}$ as in these
figures, there is a slowly rising part on the left, a flat part in
the middle-right, and a dip and second rising part on the far
right. The density wave in M33 appears in the left-most few points
in the azimuthal power spectra (Fig. 3), which turn up by a factor
of $\sim10$ compared to the extrapolated curves at low $k$. The
left-most point corresponds to a one-arm spiral, and the
second-left most point corresponds to a two-arm spiral. There is
no such systematic turn up in the spiral arm and interarm power
spectrum (Fig 4).

The similarity of the power spectra for the different passbands
may be surprising considering the galaxy looks smoother in red
colors. However the power spectrum only gives a relative measure
of power at each spatial frequency, whereas the visual impression
of smoothness is based mostly on the ratio of high frequency
emission to average brightness. Slight differences in the 3
passbands are visible at high $k$, where the prominence of the
stellar dip decreases toward the red.

The power spectra and intensity scans of a clipped H$\alpha$ image
of M33 taken with the same instrument are shown in Figure 5.  The
dashed line has a slope of 1, which is similar to the slope of the
$k^{5/3}$-normalized power spectrum of the optical emission in
Figure 3.  There are many small H$\alpha$ sources that act like
stars in the intensity scans, giving dips in the power spectra at
high frequency as for the optical images.

The top axes in Figures 3, 4, and 5 give the distance scale in
parsecs that corresponds to $2/k$ on the abscissa.  The outer
scale for the horizontal part of the power spectrum is
$2/0.02\sim100$ pixels, which corresponds to 170 parsecs. This
scale is about the same as what we found for galaxies in Paper I,
although some aspects of the overall shape result from pixelation,
as shown in the next section, and are therefore
distance-dependent.

\section{Model Power Spectra}

Fractal Brownian motion models were made by filling half of a
$640\times640\times640$ cube in $k$-space with random complex
numbers having real and imaginary values between 0 and 1, and then
multiplying these values by $k^{-\beta/2}$ for
$k=\left(k_x^2+k_y^2+k_z^2\right)^{1/2}$.  The cube is only half
full because of the symmetries in the Fourier transforms.  The
inverse Fourier transform of this cube, from complex to real
numbers, gives a three-dimensional fractal with positive and
negative numbers having a Gaussian distribution of values.  This
fractal is then exponentiated to give another fractal, now with a
log-normal distribution of all-positive values.  This last step is
done to mimic models of turbulence which produce a log-normal
distribution of density (e.g., Wada \& Norman 2001). An example of
a fractal cube made this way is in Figure 2 of Elmegreen (2002b).
Three dimensional power spectra of the model reproduce the input
power spectrum slope $\beta$ (which is twice the input power of
$k$ because the power spectrum is made from the square of the
Fourier transform), and one-dimensional power spectra of the model
have a slope that is shallower by 2. We denote the slope of the 1D
model power spectrum by $\alpha$ and confirmed experimentally that
$\alpha=\beta-2$ for $\beta$ greater than 2.  Input $\beta<2$ give
power spectra indistinguishable from noise ($\alpha=0$).  Other
fractal Brownian motion models of interstellar structure 
and a more detailed discussion of $\beta$
are in St\"utzki et al. (1998).

We varied the intrinsic power $\beta/2$ from $-0.33$ to $+2.66$
and plotted the power spectrum for each model using density scans
along a strip and along a circle in the midplane. We also
integrated the cube through one dimension with a Gaussian
weighting factor centered on the midplane, to simulate the average
Gaussian profile of the interstellar medium viewed through a disk
face-on.  These power spectra should represent the observations of
a fractal gas, and in fact models of the HI emission from the LMC
were fitted in this way (Elmegreen, Kim \& Staveley-Smith 2001).

The optical power spectra presented here are for starlight, and
models of this require an extra step.  To simulate stars, we use
small points of ``emission'' with Gaussian profiles having a
$\sigma$ of 2 pixels and an intensity equal to a random number
between 0 and 30; these values were chosen to match the width and
depth of the high-$k$ dip.  We consider three cases: (1) an image made
only of stars with a random distribution in space; (2) an image
made only of stars with a fractal distribution in space, and (3)
an image made of the fractal continuum in addition to stars with
the same fractal distribution in space. The fractal distribution
of stellar points is done by going to each cell in the fractal
model cube and choosing a random number between 0 and 1. If the
density in the cube at this point (differenced from the minimum
density and divided by the maximum density for normalization)
exceeds some factor times the random number, then we place a star
there with the assumed Gaussian profile. If the density is less
than this value, then we do not place a star there. If adjacent
points have stars, then we add all the stellar profiles together.
The multiplicative factor for the (0,1) random number determines
how many stars there are.  A value of 0.4 was chosen to give
intensity profiles that looked reasonably dense in resolved
stellar sources. In case (3), we chose stars in the same way but
added this starlight to the fractal continuum in each cell.

The most reasonable case is (3). In a typical galaxy there will be
both isolated bright stars and a haze of blended faint stars. Case
(3) assumes that both follow the fractal distribution of the gas
as a result of moderately recent star formation. The underlying
smooth distribution of old stars will not contribute much to the
high spatial frequencies but will appear mostly in the lowest few
$k$ values, like the spiral density wave.  These can be ignored in
both the modelled and the observed power spectra.

Figures 6, 7, and 8 show the results for three values of $\alpha$
which bracket our observations.  For the best-fit $\alpha=0.66$
case in Figure 7, the power spectrum and intensity scan from the
4th ellipse out in the R band image is shown at the top. The left
panel shows power spectra multiplied by $k^{5/3}$, the middle
panel shows conventional power spectra with no multiplier, and the
right panel shows the corresponding intensity scans. The power
spectrum below the R band observation is for a strip from left to
right in the fractal cube. Below that is the power spectrum from a
circle at the maximum radius of the model cube, giving a
circumference 1700 pixels long. Next comes the power spectrum from
continuum plus stars in a fractal pattern, and below that is the
power spectrum from the pure stellar image with a fractal pattern.
At the bottom is the spectrum for a pure-star image with a random
distribution of stellar positions. This pure-star power spectrum
is the same in Figures 6, 7, and 8 because it does not depend on
$\alpha$.

The best match between the star+continuum power spectrum and the
observation of M33 is for $\alpha\sim0.66$. The $\alpha=0$ case
has a power spectrum that is slightly too steep in its rising part
on the left (meaning that in a conventional plot, not normalized
with multiplication by $k^{5/3}$, the power spectrum is too
shallow in its decline here). The $\alpha=1.33$ case has a power
spectrum that is too slowly rising in this part (too steeply
falling in a conventional plot). Generally, the star+continuum
case is a better match than the fractal-star case, which has too
large a dip at high frequency from the intrinsic stellar profiles.
The random star case is a poor fit, having a power spectrum in the
normalized plot with an increase as $k^{5/3}$, which corresponds
to a flat power spectrum in a conventional plot (as seen at the
bottom of the middle panel of Fig. 7).

The azimuthal power spectrum from the fifth ellipse out in the
I-band image of NGC 5055 (Paper I; data from HST archives) is
reproduced in Figure 9, along with several models. This galaxy has
fewer foreground stars and so the dip at high frequency is very
slight. The two models at the bottom are pure stars in a fractal
pattern. The bottom one has a star brightness that is a random
number multiplied by 5 and the next one up has a star brightness
that is a random number multiplied by 20.  Both star
models have Gaussian $\sigma=2$ px. The power spectra are
identical although the intensity traces on the right show these
differences clearly. Next up from the bottom are three models with
a fractal distribution of stars and the associated continuum. They
each have a star brightness equal to a random number times 5 and
the stars have $\sigma=2$ px again. This
is only one-sixth the star brightness used for the M33 models, and
the high-frequency dip is smaller too, in proportion. The three
models differ in the slope of the intrinsic power spectrum,
decreasing from $\alpha=1.70$ to $1.50$ to $1.33$ with increasing
height in the figure.  The power spectra show this difference at
low frequency where the rising part gets slightly steeper as
$\alpha$ decreases. The middle one, with $\alpha=1.50$, is the
best fit to NGC 5055. The power spectrum of the pure continuum
with no stars in this case is shown next to the observation of NGC
5055. It differs from the star+continuum case only slightly at
high frequency. We conclude from these models that NGC 5055 has
relatively faint point sources and an intrinsic power spectrum for
the optical emission with a slope of $-1.5\pm0.2$.

The $k^{-1.5\pm0.2}$ power spectrum fit to NGC 5055 is clearly
steeper than the $k^{-0.66\pm0.66}$ fit to M33. The difference
could be the result of residual stars and unresolved clusters in
the M33 galaxy, although NGC 5055, which is 9 times more distant,
has 4 times better resolution from HST so the pixel scale in
parsecs (3.5 pc) is only a factor of 2 greater for NGC 5055 than
for M33. Residual foreground stars could be more important. M33
has many more foreground stars than NGC 5055 because of its large
angular size. Foreground stars are randomly positioned and their
power spectrum varies like $k^0$, as in Figures 6, 7, and 8.
Contamination from numerous random stars will flatten the
spectrum.

Figure 10 shows a fit to M33 that starts with the underlying galaxy
model for NGC 5055 and adds random foreground stars with an amplitude
6 times larger than in the NGC 5055 model itself and
a stellar width of $\sigma=2$ px.  At the bottom is the
power spectrum and intensity scan for the foreground stars. Next is the
observation of NGC 5055, from Figure 9. The third curves up are the power
spectrum and intensity scan for the sum of the best-fit model of NGC 5055
(the $\alpha=1.5$ model in Fig. 9) and the foreground stars.  The top
curves are the observations of M33.  The model of M33 with foreground
stars has the same intrinsic galactic properties as NGC 5055, namely a
superposition of galactic stars and a continuum that both follow a fractal
pattern giving a 1D power spectrum slope of $-1.5$.  The foreground stars
make the slope appear shallower, $-0.66$, like the observed slope in M33.

\section{Conclusions}

The power spectra of optical emission in galaxies show a characteristic
structure that can be modelled by a superposition of a fractal unresolved
brightness distribution and a corresponding fractal distribution of point-like
sources. The intrinsic power spectrum of these distributions has a slope
of $-0.66\pm0.66$ for M33 and a slope of $-1.5\pm0.2$ for NGC 5055. The
relatively flat slope for M33 is probably the result of foreground
stars, which are difficult to remove from the observations.  
A model of M33 with
foreground stars superposed on the NGC 5055 model demonstrates this point.
The power spectrum slope for the foreground-corrected version of M33,
and the slopes for NGC 5055 and the other galaxies in Paper I are
all comparable to the one-dimensional power spectrum from
turbulence. We infer from this that young stars and clusters are
distributed in a fractal pattern that is generated by a turbulent gas.

Acknowledgements:
B.G.E is supported by NSF grant AST-0205097.

\newpage
\begin{figure}
\caption{R-band images of
M33 with the azimuthal and spiral strips
used for power spectrum analysis.
The widths of the
strips are 9 and 21 pixels, respectively.
North is up.
Power spectra were made from single-pixel wide strips
and then averaged together.}
\end{figure}

\begin{figure}
\caption{Intensity scan for the fourth ellipse
out from the center of the R-band image of M33 (top) and the power
spectrum multiplied by $k^{5/3}$ of this scan (bottom).  The
original scan and power spectrum are on the left, and the version
with the brightest stars removed is on the right. The dashed line
on the right indicates the extent of the horizontal part in the
power spectrum.}
\end{figure}

\begin{figure}
\caption{Power spectra multiplied by $k^{5/3}$ of
ten radii in each of 3 passbands for M33. The radii are indicated
in figure 1; here the radius increases toward the top of the
plot.}
\end{figure}

\begin{figure}
\caption{Power spectra multiplied by $k^{5/3}$
and intensity scans for 8 spiral and interspiral traces in the R
band image of M33. The exponential disk is evident in the
intensity scans. The ends are tapered with a cosine function to
prevent ringing in the power spectrum. Intensity peaks in this
image can easily be correlated with bright patches along the
spiral lines in Figure 1. The bottom intensity scan and
corresponding power spectrum in this figure correspond to the southern
spiral strip that goes through the main spiral arm at the
bottom of Figure 1.
The other scans going up here follow in clockwise order in Figure
1. The northern main spiral arm is the 5th up from the bottom here.
The arm scans show the brightest emission in the intensity
profiles and have the flattest power spectra at long wavelengths.}
\end{figure}

\begin{figure}
\caption{Power spectra multiplied by $k^{5/3}$
and intensity scans of a clipped H$\alpha$ image of M33. The
dashed line at the top of the left panel has a slope of 1. }
\end{figure}

\begin{figure}
\caption{Model results for a fractal cube made
with an intrinsic 1D power spectrum having a power law slope of
$0$. The bottom power spectrum on the left and the bottom
intensity scan on the right are for a random distribution of stars
unrelated to the fractal. The dashed line near the power spectrum
has a slope of $5/3$, meaning that the un-normalized power
spectrum is flat. The second curves up from the bottom are for
stars arranged in a fractal pattern. They are based on azimuthal
profiles through the midplane of a 640-cubed fractal density
distribution, placing a star everywhere the density of the
continuum exceeds the peak density value multiplied by a random
number that is uniformly distributed between 0 and 0.4.  The third
scan up from the bottom is this same fractal star distribution
added to the fractal continuum distribution. The fourth scan is
the continuum only, and the fifth scan is the continuum only in a
linear scan through the midplane. }
\end{figure}

\begin{figure}
 \caption{Model results for a fractal cube made
with an intrinsic 1D power spectrum having a power law slope of
$-0.66$. The bottom 5 curves are as in Fig. 5, and the top curves
in each panel are the star-reduced R band result for the 4th
ellipse out from the center of M33.  The panel on the left
contains the power spectra multiplied by $k^{5/3}$, as in the
other figures here; the panel in the middle has the original power
spectra without the normalization.  This case with $\alpha=0.66$
is the best fit to M33 for the model with a fractal continuum
and stars (third curve up from the bottom in each panel).
The top
panel on the right is drawn separately to
contain the longer scan length for M33.}
\end{figure}

\begin{figure}
\caption{Model results for a fractal cube made
with an intrinsic 1D power spectrum having a power law slope of
$-1.33$. }
\end{figure}

\begin{figure}
\caption{The power spectrum of the fifth
intensity scan from the center of the I-band image of NGC 5055,
from paper I, is shown at the top in this figure, along with the
intensity scan itself on the right, and models that bracket the
preferred fit are shown below.  }
\end{figure}

\begin{figure}
\caption{Another model of M33
is shown here, third up from the bottom. This was made
from the best-fit model of NGC 5055
by adding a Poisson distribution of foreground stars. 
The bottom scans here are the power spectrum and
intensity scan for these foreground stars. 
The second and fourth scans up from the bottom
are the observations of NGC 5055 and M33. 
Dashed lines show matching slopes. The top
panel on the right is drawn separately to
contain the longer scan length for M33.
}
\end{figure}

\end{document}